\documentclass{sig-alternate-2013}

\usepackage{wasysym}
\usepackage{balance}
\usepackage{booktabs} 
\usepackage{cite}
\usepackage{color}
\usepackage{algorithm}
\usepackage{algorithmic}
\usepackage{stackrel}
\usepackage{siunitx}
\usepackage{xspace}   
\usepackage{enumitem} 
\usepackage{url}

\newcommand{\Paragraph}[1]{\smallskip\noindent{\bf #1.}}

\newcommand{\eg}{e.\@\,g.\@\xspace}

\newcommand{\Tdot}{$\CIRCLE$}

\newcommand{\Twdot}{$\Circle$}

\def\sutdlong{Singapore University of Technology and Design\xspace}
\def\swat{SWaT}
\def\swatlong{Secure Water Treatment}

\makeatletter
\def\@copyrightspace{\relax}
 \makeatother

\begin{document}
\include{license}

\title{On Ladder Logic Bombs in Industrial Control Systems}

\numberofauthors{3}
\author{
\alignauthor Naman Govil\\
\email{naman.govil\\@research.iiit.ac.in} \\
\affaddr{IIIT Hyderabad}\\
\affaddr{Hyderabad - 500032}\\
\affaddr{India}\\
\alignauthor Anand Agrawal \\
\email{agrawal\_anand\\@sutd.edu.sg}\\
\affaddr{Singapore University of Technology and Design}\\
 \affaddr{487372 Singapore}\\
\alignauthor Nils Ole Tippenhauer\\
\email{nils\_tippenhauer\\@sutd.edu.sg}\\
\affaddr{Singapore University of Technology and Design}\\
 \affaddr{487372 Singapore}\\
}


\maketitle
\begin{abstract}
  In industrial control systems, devices such as Programmable Logic
  Controllers (PLCs) are commonly used to directly interact with
  sensors and actuators, and perform local automatic control. PLCs run
  software on two different layers: a) firmware (i.e. the OS) and b)
  control logic (processing sensor readings to determine control
  actions). 

  In this work, we discuss \emph{ladder logic bombs},
  i.e. malware written in ladder logic (or one of the other IEC
  61131-3-compatible languages). Such malware would be inserted by an
  attacker into existing control logic on a PLC, and either
  persistently change the behavior, or wait for specific trigger
  signals to activate malicious behaviour. For example, 
  the LLB
  could replace legitimate sensor readings 
  with manipulated values. We see the
  concept of LLBs as a generalization of attacks such as the Stuxnet
  attack. We introduce LLBs on an abstract level, and then demonstrate
  several designs based on real PLC devices in our lab. In particular,
  we also focus on \emph{stealthy LLBs}, i.e. LLBs that are hard to
  detect by human operators manually validating the program running in
  PLCs.

  In addition to introducing vulnerabilities on the logic layer, 
  we also discuss countermeasures and we propose two detection techniques.
\end{abstract}


\section{Introduction}
Industrial Control Systems (ICS) are computer systems that typically
control physical processes that relate to power, water, gas,
manufacturing and other critical infrastructure. ICS and Supervisory
Control and Data Acquisition (SCADA) systems rely on local
programmable logic controllers (PLCs) to interface with sensors and
actuators. While PLC devices are available from a range of
manufacturers, they are all commonly programmed with the same set of
programming languages based on IEC 61131-3. In particular, the IEC
61131-3 standard~\cite{john10iec611313} contains \emph{ladder logic},
\emph{functional block diagram}, and \emph{sequential text} as
different languages that are used together to define logic to run on
the PLCs. The logic is then interpreted by the firmware running on the
PLCs. Modern PLCs provide security mechanisms to allow only legitimate
(e.g., signed) firmware to be uploaded. In contrast, logic running on
the PLCs can typically be altered by anyone with network or local USB
access to the PLC. This setting is the main difference to malware
scenarios in traditional corporate IT environments, where the
injection of attacker code is usually significantly harder.

Recently, the security of Cyber Physical Systems (CPS) and related
systems has gained a lot of attention~\cite{chabukswar2010simulation,
  LinYangXuZhao, wangYummingXiaofeiYiuHuiChow, zhuJosephSastry,
  zonouzRogersBerthierBobbaSandersOverbye}. In particular, CPS such as
critical infrastructure including power grids, nuclear power plants,
and chemical plants are threatened. In CPS, physical-layer
interactions between components have to be considered as potential
attack vectors, in addition to the conventional network-based attacks.

In this work, we introduce \emph{ladder logic bombs} (LLBs),
i.e. malware written in ladder logic (or one of the other IEC
61131-3-compatible languages). LLBs consist of logic that is intended
to disrupt the normal operations of a PLC by either persistently
changing the behaviour, or by waiting for specific trigger signals to
activate malicious behaviour. In particular, the LLBs could lay
dormant and hence hidden for a very long time until a specific trigger
is observed.  Once activated, the LLB could
replace legitimate sensor readings that are being reported by the PLC
to the SCADA system with manipulated values. We introduce LLBs by
classifying their purpose and action, and demonstrate several
constructions based on real PLC devices in our lab.  

We implemented and tested our attacks on a real-world ICS 
(the \swat testbed, see Section~\ref{sec:implementation}). In particular, we
focused on \emph{stealthy LLBs}, i.e. LLBs that are hard to detect by
human operators manually validating the program running in PLCs. We
provide a classification of logic based attacks, such as the ones
performed by Stuxnet~\cite{falliere2011w32}.  


We summarize our contributions as following:
  \begin{itemize}[noitemsep,nolistsep]
  \item We analyzed firmware updates on the target platform to detect
    vulnerabilities.
  \item We identify the issue of logic manipulations on PLCs, and
    introduce the concept of \emph{ladder logic bombs} (LLBs).
  \item We present a range of LLB prototypes, in particular ones that 
  attempt to hide from manual logic code inspection.
  \item We discuss countermeasures based on manual and automatic code 
  inspection, and a central-server based solution.
  \end{itemize}

  The structure of this work is as follows: In
  Section~\ref{sec:background}, we introduce CPS systems, PLCs, and
  IEC 61131-3 in general. We propose our Ladder Logic Bomb concept in
  Section~\ref{sec:bombs}, and present example implementations in
  Section~\ref{sec:implementation}. 
  The results of a small-scale evaluation are summarized in Section~\ref{sec:evaluation}. 
  We propose a countermeasure against LLB attacks in
  Section~\ref{sec:countermeasures}. Related work
  is summarized in Section~\ref{sec:related}. We conclude the paper in
  Section~\ref{sec:conclusions}.

\section{Background}
\label{sec:background}
In this section, we will introduce some of the salient properties
of industrial control system (ICS) networks that we have found so far. In
addition, we will briefly introduce Ladder Logic programming language and
the tools necessary to interact with such PLCs.

\subsection{ICS}

In the context of this work, we consider ICS that are used to
supervise and control system like public infrastructure (water,
power), manufacturing lines, or public transportation systems. In
particular, we assume the system consists of programmable logic
controllers, sensors, actuators, and supervisory components such as
human-machine interfaces and servers. We focus on single-site systems
with local connections, long distance connections would in addition
require components such as remote terminal units (see below). All
these components are connected through a common network topology.

\Paragraph{Programmable logic controllers} PLCs are directly
controlling parts of the system by aggregating sensor readings, and
following their control logic to produce commands for connected
actuators.

\Paragraph{Sensors and actuators} Those components interact with the
physical layer, and are directly connected to the Ethernet network (or
indirectly via remote input/output units (IOs) or PLCs).

\Paragraph{Network Devices} ICS often use \emph{gateway} devices to
translate between different industrial protocols (\eg Modbus/TCP and
Modbus/RTU) or communication media. In the case where these gateways
connect to a WAN, they are usually called \emph{remote terminal units}
(RTUs).

\begin{figure}[tb]
\centering
\includegraphics[width=\linewidth]{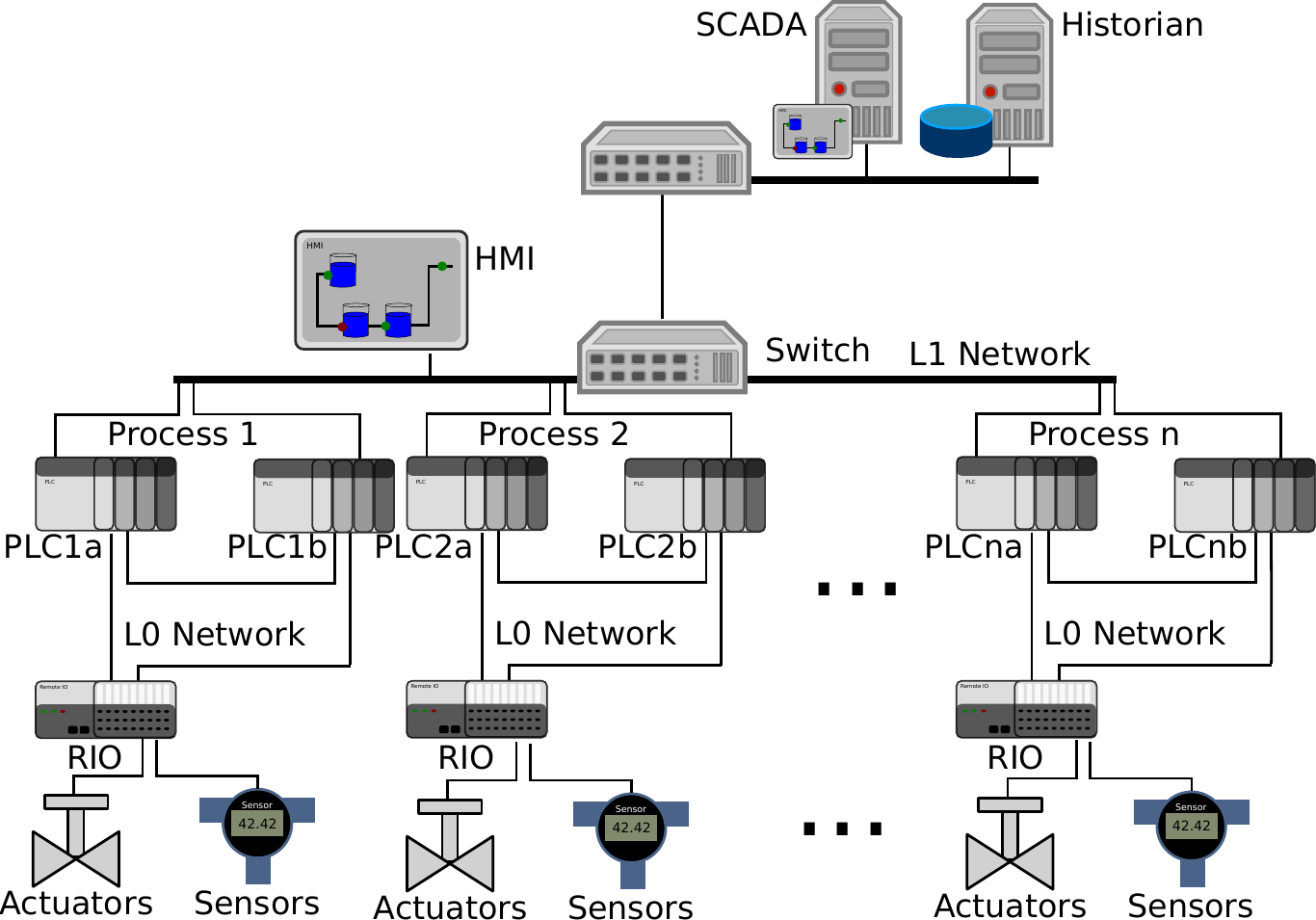}
\caption{Example local network topology of a plant control network.}
 \label{fig:generic}
\end{figure}

\subsection{Ladder Logic and Studio 5000}
A Programmable Logic Controller (PLC) is an industrial computer system
that continuously monitors the state of input devices and makes
decisions based on a custom program to control the state of output
devices. PLCs are widely used in industrial control systems (ICS) for
handling sensors and actuators, mainly because of their robust nature
and ability to withstand harsh conditions including severe heat, cold,
dust, and extreme moisture.  Considering their widespread usage and
important nature of tasks handled by PLCs, their security against
malicious manipulation is critical.

PLCs are user programmable devices. PLC programs are typically written
in a special application on a local host (personal computer), and then
downloaded by either a direct-connection cable or over a network to
the PLC. The program is stored in the PLC in a non-volatile flash
memory. While details differ for platforms from alternative vendors,
it might be required to enable remote change of control software on
the PLC through a physical switch (i.e., \emph{program} mode on
ControlLogix devices). We observe that due to convenience, in
practical systems PLCs are often kept in that setting to allow easy
remote access. In addition, any attacker with physical access is able
to change the switch setting easily. For that reason, we assume that
remote or local reprogramming access is possible in the
remainder of this work.

IEC 61131-3 is an open international standard~\cite{john10iec611313} for PLCs that defines 
2 Graphical and 1 Textual programming language standards for PLCs:
\begin{itemize}[noitemsep,nolistsep]
\item Ladder Logic Diagrams(graphical)
\item Functional block Diagram (graphical)
\item Structured Text (textual)
\end{itemize}
  
\begin{figure}[tb]
\centering
\includegraphics[width=\linewidth]{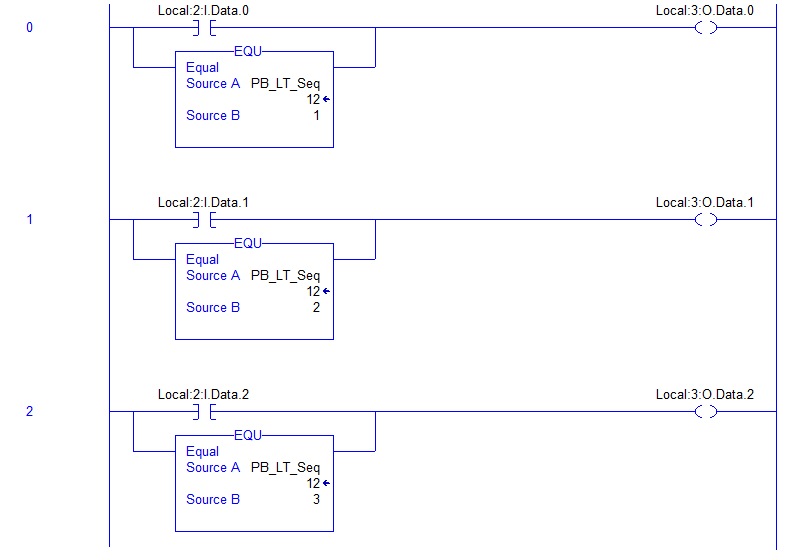}
\caption{Example ladder logic code with three rungs}
 \label{fig:intro}
\end{figure}

The most popular of those languages is Ladder Logic Diagrams. The main
intuition behind this Ladder Logic Diagrams is to provide a system-wiring
diagram abstraction similar to electro-mechanical relays. Ladder
logic is more of a rule-based graphical language implemented by
\emph{rungs}, rather than traditional procedural-based language. A
rung in the ladder represents a rule. They are called ``ladder''
diagrams because they resemble a ladder, with two vertical rails
(supply power) and as many "rungs" (horizontal lines) as there are
control circuits to represent.  Figure~\ref{fig:intro} depicts an
example logic implemented in ladder logic diagram. It contains three
rungs which utilize various inputs, outputs and instruction blocks (If
Equal To block here) to implement certain logic.

"Studio 5000" is a software product of Rockwell Automation that
provides an environment to develop a range of elements for a control
system, for operational and maintenance use. Its major element is the Studio
5000 Logix Designer application, formerly (RSLogix 5000), software to
program Logix5000 controllers.
 
Another tool called RSLinx is used to establish USB-based
communication between PLCs and a host PC running Studio 5000. RSLinx is
a Windows based software package to interface with a range of ICS and
automation hardware.  In this paper, we used Allan-Bradley PLCs
(ControlLogix 5571) with Studio 5000 v21.00. It is important to note
that for different PLCs, different versions of RSLinx and Studio 5000
have to be used.

\subsection{Analysis of PLC Vulnerabilities}
As part of our investigations for this work, we carried out an analysis
to explore vulnerabilities in the firmware running on PLCs such as
the ControlLogix 5571. We briefly summarize our results here.

We investigated whether a local attacker with physical access to the
PLC (or remote access via network) would be able to a) obtain the
currently running firmware from the PLC, and b) upload a modified
version of the firmware. To put this into perspective, the latter
technique could be used to install hidden backdoors/trojans in the
firmware, and/or to change other operational behavior of the
firmware. We found that using a local USB connection, we were always
able to obtain the running firmware. In addition, it was possible to
obtain the firmware via the network, if the PLC was set into
\emph{programming} mode via a hardware selector switch. In the
following, we discuss the second action, the upload of modified
firmware.

The firmware for our PLC devices is distributed as a \emph{.dmk}
file. This file contains two sets of binary files (\emph{.bin}) and
associated digital certificates (\emph{.der}). It also contains a
\emph{.nvs} file which includes information about the firmware
version, product code/type, etc. It is this file which acts as a
header file for all the other files, linking each binary image with
its respective certificate and also mentions the load address for
every file.

\begin{figure}[tb]
\centering
\includegraphics[width=\linewidth]{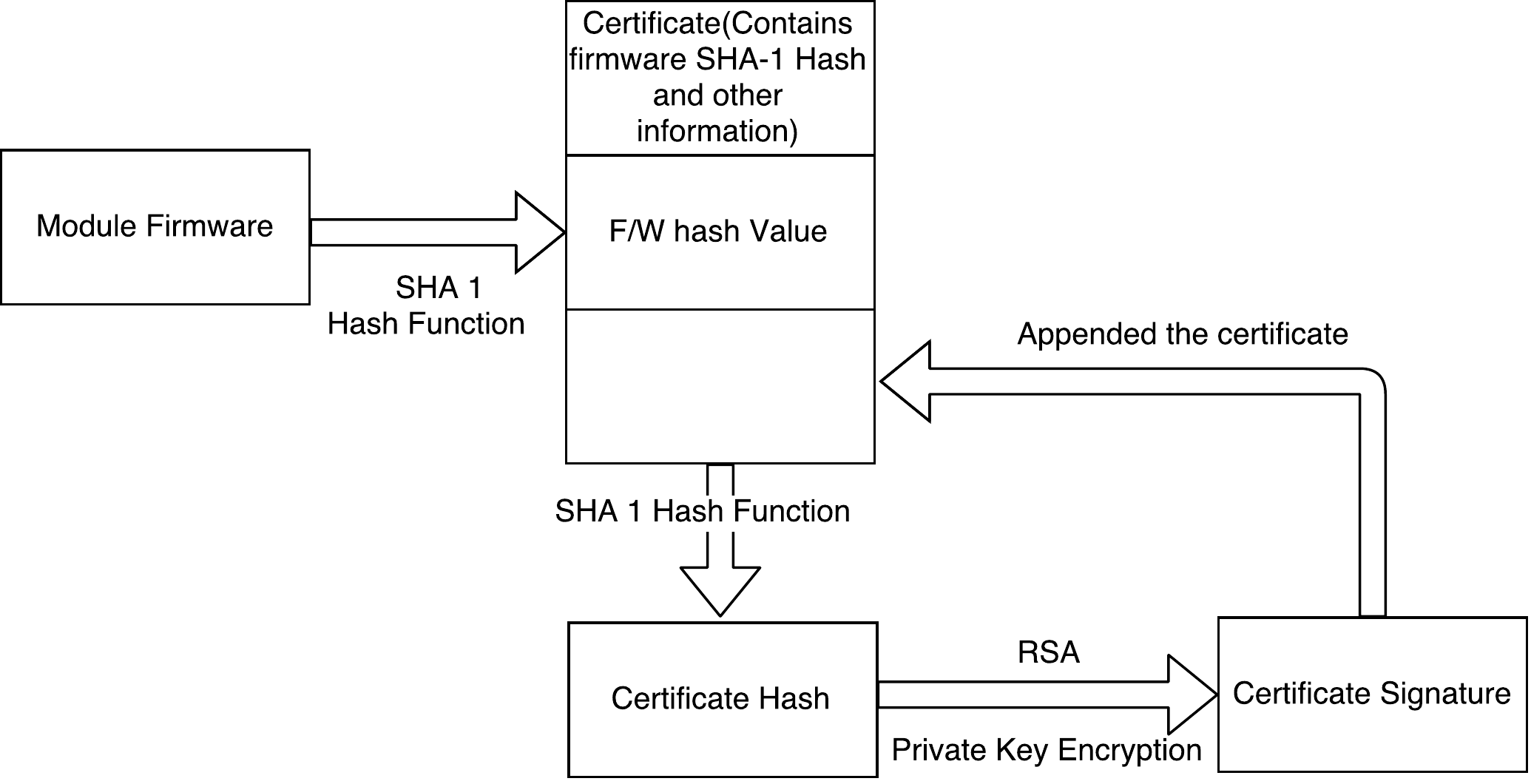}
\caption{Firmware Signing and Authentication Process~\cite{batke2013methods}}
 \label{fig:firmwareauth}
\end{figure}

The digital certificate is signed by the manufacturer (Rockwell
Automation). This digital signature is the hash value of the
certificate itself, encrypted with RSA algorithm using the private key
of the manufacturer. In addition, the certificates also contain a
cryptographic hashsum (using SHA-1) of the firmware image in one of
its data fields. At firmware update time, the module (PLC) receives
the certificate containing the firmware's hash value and the
certificate's digital signature. The module computes the hash value of
the certificate, decrypts the signature, and compares the hash
values. If these hash values match, then the certificate is valid.  If
not, the update is rejected. After receiving the entire firmware
image, the module then computes the hash value of the firmware and
compares it with the value from the certificate. If the values do not
match, the firmware update is rejected. Given this construction, any
modifications to the firmware image by the attacker will change the
hash sum, leading to a mis-match between the hash sum already existing
in the certificate. Any change of the hash sum in the certificate will
invalidate the signature by the manufacturer. This process is
explained in more detail in Figure~\ref{fig:firmwareauth}

\begin{figure}[tb]
\centering
\includegraphics[width=\linewidth]{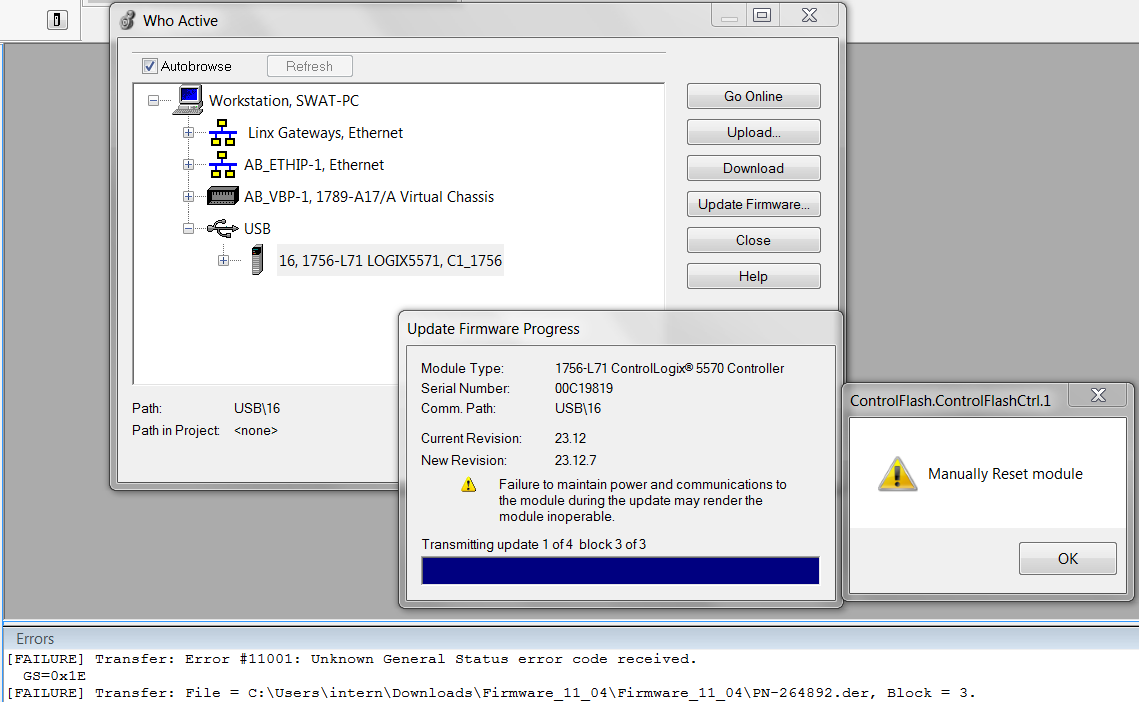}
\caption{Flashing faulty firmware (with modified certificate)}
 \label{fig:certificate}
\end{figure}

Given the described setup, we decided to check whether the
certificates were correctly validated by the PLC. We removed the
original certificate of a valid firmware, and replaced it with a
certificate that was signed by our own (self-signed) CA. We ensured
that the spoofed certificate had matching content in every custom data
fields, to match the valid firmware image and its hash value.  Then,
we used the resulting file (our own \emph{.der}) to update the
firmware on a PLC. The update failed, and we received an error
(Transfer: Error \#11001). The process can be found in
Figure~\ref{fig:certificate}, where it can be seen that the error is
triggered when trying to upload the custom certificate.
As conclusion, we currently assume that the firmware update mechanism
is sufficiently secured against manipulations by an attacker.  As a
next step, we evaluated the PLC logic update process. We discovered
that there were absolutely no checks/verifications performed to ensure
that logic updates being pushed onto the PLC are coming from
authorized sources. 
In the following, we concentrate on such manipulations of the PLC
logic.

\section{Ladder Logic Bombs}
\label{sec:bombs}
In this section, we present our proposed concept of \emph{ladder logic
  bombs}.  In particular, we noticed that while changes to the
firmware of PLCs are made more difficult by digital signatures, the
actual logic that is executed on the PLCs is not protected by such a
measure. In addition, the lack of security checks/authentication
before downloading new logic onto PLCs is a cause of major concern. An
attacker can exploit this by either gaining physical access to the
PLCs or over the network, and can download custom (\emph{malicious})
logic onto PLCs which can compromise the system.  Next, we discuss
potential attack scenarios and goals, which can be achieved through
this vulnerability.
\subsection{System and Attacker Model}
\label{sec:attackerModel}
In this work, we assume that the attacker is able to access PLCs in an
industrial control system either remotely via the network, or
physically. As we will show, commonly such access will allow the
attacker to read and modify the programming logic of the PLCs without
any authentication. The attacker is assumed to have access to the
respective software required to download and upload logic
configurations to the PLC (e.g., Studio 5000 for ControlLogix PLCs).

The goals of the attacker can range from achieving a Denial of Service
(DoS), to changing the behavior of the PLCs, or to obtain data traces
of sensor and control messages processed by the PLC. In order to
perform these attacks, the attacker just needs access to the PLC
system once, making such attacks all the more dangerous.  The attacker
could also have sporadic (physical) access to the PLC. For example,
the attacker only has access to the PLC once a week (because he is a
regular contractor). In these events, the attacker can trigger any
behavior changes (i.e trigger his ladder logic bomb) at a point
unrelated to his access time (e.g., to hide correlations to his
access).

The system we consider in this setting is very generic and can be
described as follows: a PLC in an industrial control system which uses
IEC 61131-3 languages for the logic, and can be re-programmed as
described above. It is connected to sensors and actuators of a
critical process. Operators of the plant configured the logic of the
PLC at design time. Though they continuously monitor the status of
these PLCs, they seldom need to change the logic configuration of the
already operational system. They are also able to manually download
the logic to inspect it, if required. Although we will briefly discuss
a network-based detection mechanism using an intrusion detection
system later, such a solution will not be able to detect changes by a
local attacker. For that reason, we do not focus on IDS in this
work. In addition, physical layer prevention mechanisms (camera,
fences, etc) are out of scope of this work.

We do not consider an attacker that is able to attack the operator's
machines (as it was the case in Stuxnet), or able to manipulate
network traffic while it is being transmitted.  In particular, if the
attacker was able to compromise the operator's machine, then the
operator would not be able to verify any code reliably. Such an
attacker could be addressed by using a trusted computing platform,
which we consider out of scope for this work. The attacker model does
also not consider insider attacks (e.g., an attacker who might be
regular contractor/employee with authorization to access and modify
the PLC logic).

\subsection{Bomb Classification}
\begin{figure}[tb]
\centering
\includegraphics[width=\linewidth]{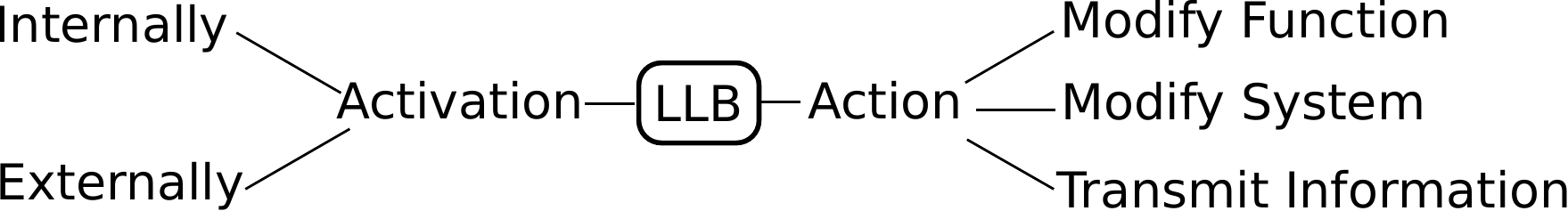}
\caption{Ladder Logic Bomb (LLB) Classification}
 \label{fig:llbclassification}
\end{figure}

Ladder logic bombs can be classified broadly by two criteria (as shown
in Figure~\ref{fig:llbclassification}).
LLBs can be classified according to their activation and triggering.
They can either be externally triggered by giving a certain
input. Alternatively, they can be triggered by internal logic (system
states, specific instructions or data, clock, etc.)

LLBs can also be classified according to the alteration they incur onto 
the existing PLC system. They can add or remove certain functionality in 
the existing logic (\emph{modify function}). These bombs can also alter 
the system values such as system date/time, timezone, wall-clock time, or similar (\emph{modify system}). Finally, these can also be used for data 
exfiltration and transmitting crucial system data to a spy node (\emph{transmit information}).  

Together, those classifications now describe more specific LLBs. For
example, a LLB that turns off a pump at 12 AM would be classified as
internally activated function modification LLB. 

\subsection{Payload Types}
In the following, we present a range of payload types, that can be
used to achieve the attacker's goals as outlined in
Section~\ref{sec:attackerModel}. The payloads can be openly
destructive (e.g. causing a denial of service (DoS)), or enable
stealthy attacks (e.g., by establishing Man-in-the-Middle
capabilities). The MitM payload can be used to either eavesdrop on
traffic passing through the nodes, or potentially manipulate the
content of those messages undetected. By manipulating the message
content, the attacker can falsify sensor readings reported to other
PLCs and the SCADA system, or change commands sent to actuators. In
the following, we present these attack goals in detail.
\subsubsection{Denial of Service LLBs}
A very basic (but destructive) payload performs a Denial of Service
(DoS) attack on the PLCs.  By adding a malicious piece of logic,
hidden in the entire ladder logic of a certain PLC, which is triggered
at a specific instant can throw the PLC off control and cause it to
halt.  This could damage the process being controlled by the PLC and
could potentially cause a performance-threatening state in the system.
Such a bomb would continuously be looking for the trigger condition,
and as soon as it is met, it could launch into an infinite-loop,
repetitive subroutine calls, etc and render the PLC useless.

\subsubsection{LLBs to Manipulate Sensor Readings and Commands}
\begin{figure}[tb]
\centering
\includegraphics[width=\linewidth]{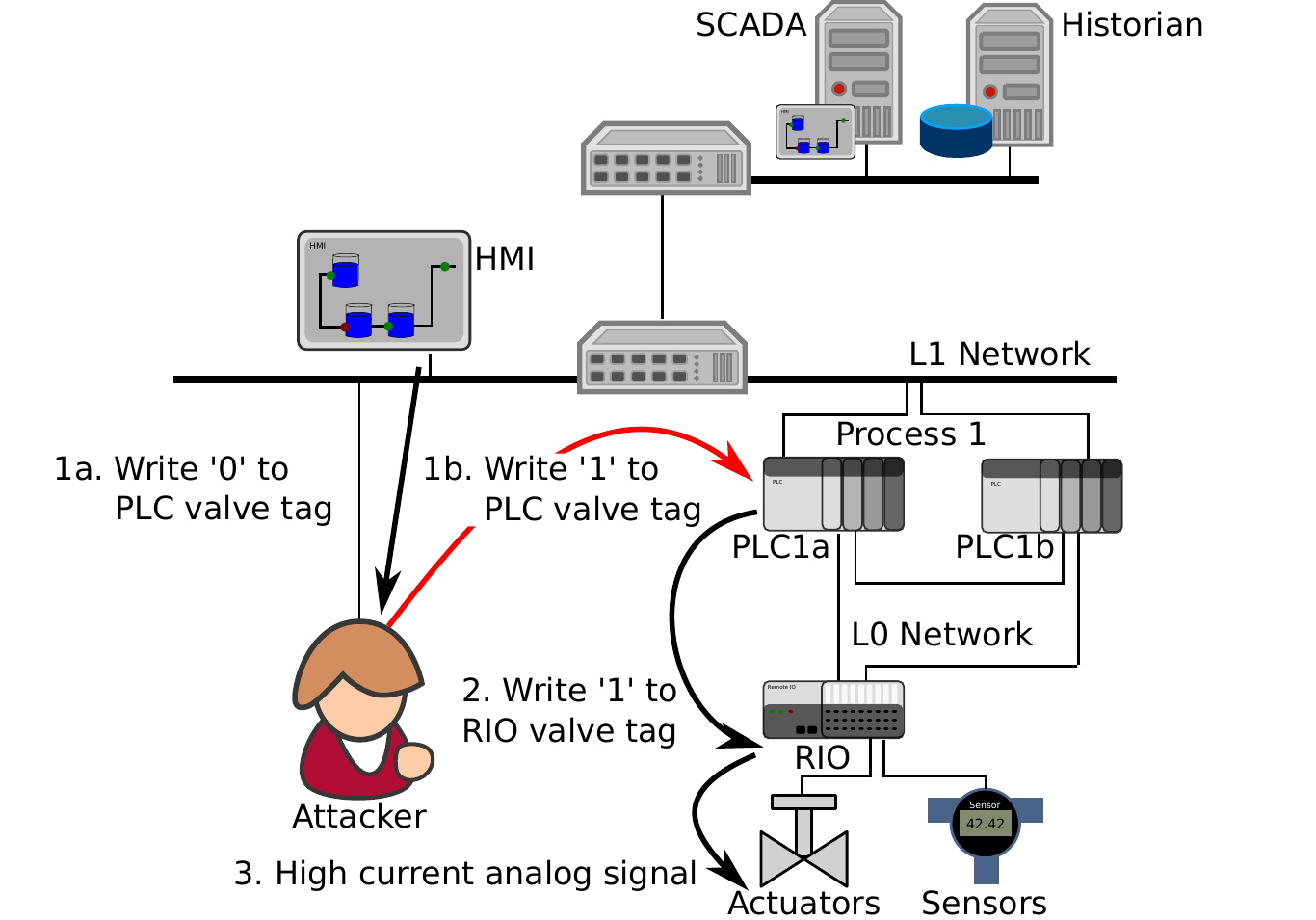} 
\caption{Manipulating sensor readings from RIO to HMI/ control instructions
	from HMI to RIO}
 \label{fig:llbmanipulate}
\end{figure}

Another class of LLBs could be used to tamper actual data being
used/generated in the PLCs. The easiest targets for such an attack are
the sensor values being read from the remote IOs (RIOs in
Figure~\ref{fig:generic}). These values could be manipulated to cause
the system to go into an unwanted state. (Figure~\ref{fig:llbmanipulate})

\subsubsection{LLBs for Stealthy Data Logging}
A third category of LLBs could be used to secretly track and keep a
log of sensitive PLC data. This can be achieved through the use of
FIFO buffers and recording data into arrays on the PLCs. These kind of
bombs are particularly dangerous, as these do not disturb the working
of the system, making the host completely unaware of their
presence. These can stay within the logic for extended periods of time
without detection, constantly leaking sensitive data and commands.

\subsection{Triggering}
Here, we describe the different triggering mechanisms that can be used with ladder logic bombs. 

\paragraph{Triggering at a particular Input}
The bomb could be set off when a pre-determined input is
detected. For example, we are targeting a water treatment ICS for our
experiments (see Section~\ref{sec:testbed}). The target PLC is
receiving inputs about the water level in one of the tanks from its
corresponding level sensor.  The bomb could be set off when a
particular level is reached in the tank.
\paragraph{Triggering Sequence}
The bomb could also be triggered when a particular trigger sequence is
detected. This would potentially make the bomb more difficult to
detect, as none of its effects would be visible until the particular
sequence is detected as input. This can be achieved by implementing a
finite state machine (FSM) using latches. 
\paragraph{Timer}
The bomb could also be set off using a timer. This would make the LLB like
a real world time bomb, which sets into motion when the timer has finished
its count sequence. Using nested TON timers, it is possible to implement
count sequences which will last days.  
\paragraph{Specific Internal Condition}
The bomb could be triggered when a particular internal state is achieved.
This particular triggering scheme requires the attacker to have complete
knowledge and understanding of the logic on the PLCs. When a particular
state variable, for example a fault code, is set, the bomb could be set
off and the payload logic is executed.

\subsection{Hiding LLBs in PLC Logic}
The na\"ive approach to detect any modifications in the original logic (in
our case, the LLBs) would be to download the control logic from
PLC devices, and manually inspect them for code changes. In
particular, engineers familiar with the plant operations might be able
to read through the code and detect malicious changes. While that
approach might be feasible for small sites and very simple logic, we
will show in the following section that there are several options for the
attacker to hide the malicious payload within the logic to make it
harder to detect by such manual inspection.

\section{Implementation}
In this section, we describe in detail the construction of ladder logic
bombs and demonstrate how they can be used to disturb the functioning of
ICS. 
\label{sec:implementation}
\subsection{\swat \ Testbed}
\label{sec:testbed}
\begin{figure}[tb]
\centering
\includegraphics[width=\linewidth]{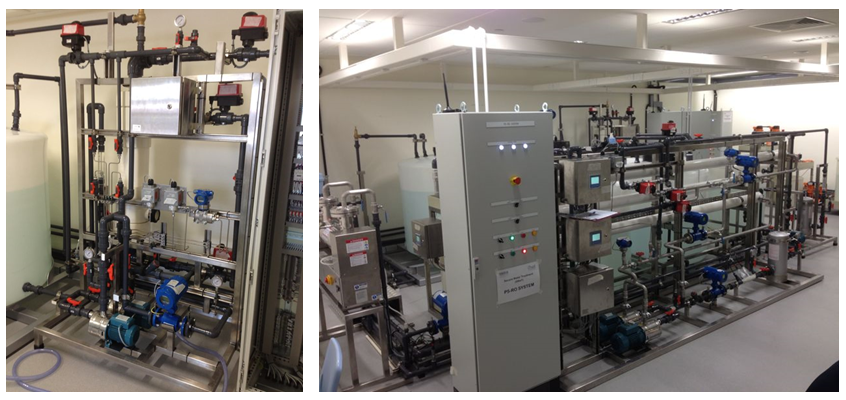}
\caption{Overview of \swat Testbed.}
 \label{fig:swat}
\end{figure}

The experiments were conducted on an industrial control system
testbed, called \swat, located at the \sutdlong. \swatlong, as
depicted in Figure~\ref{fig:swat}, is a fully functional (scaled down)
water treatment plant. \swat was constructed exclusively as platform
for research on cyber physical system security. The water treatment
process is partitioned into six stages, starting with raw water in
Tank 1 to filtered output water in Tank 6. Each stage is controlled by
an independent PLC which determines control actions using data from
sensors.

Sensors values and actuator commands are communicated to and from a PLC 
via a plant network. The system also contains monitors to view and ensure 
system states are within acceptable operational boundaries. Data from 
sensors are available for inspection on the Supervisory Control and Data 
Acquisition (SCADA) workstation and recorded by the Historian for 
subsequent analysis.

\subsection{Attack 1: DoS using Add On Instructions}
The Denial of Service (DoS) is a potential attack goal to inflict
(most often financial or reputation) damage on critical systems. In a
DoS attack, the attacker temporarily or permanently slows or stops
correct operations of a system. On the Internet, (distributed) DoS
attacks are often achieved by creating massive amounts of traffic that
overload communication links or servers. As PLCs control the action of
sensor and actuators in the system, their operational availability is
often critical~\cite{krotofil2014cps}. If the PLC is incapable of
controlling the actuators, it can have disastrous consequences (e.g.,
lead to the loss of control of heavy machinery in an automobile
assembly plant).

\textit{Goal:} In this setup, the goal was to launch a DoS attack on
one of the PLCs in a water treatment plant. 

\textit{Construction:} This has been achieved by implementing an infinite
loop as the bomb payload. The trigger mechanism for this LLB is when a
particular input is received. Similar to Stuxnet~\cite{falliere2011w32},
the trigger check condition lays on top of the actual logic, which always
stays on to check if the particular input has been received. As soon as
the desired trigger input is received, the LLB springs into action. 

\begin{figure}[tb]
\centering
\includegraphics[width=\linewidth]{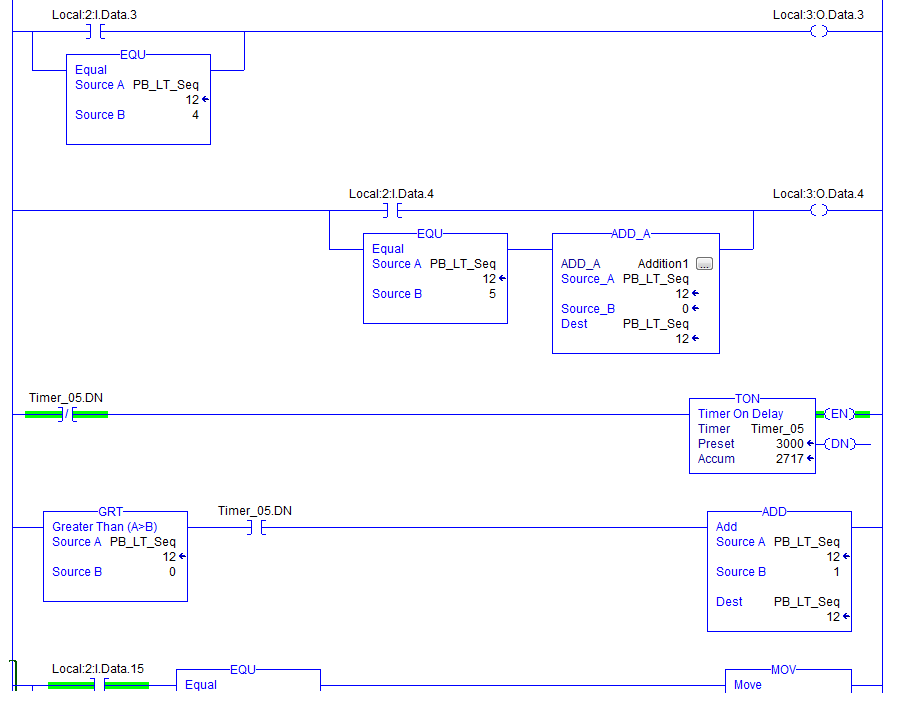}
\caption{Malicious Add-On Instruction}
 \label{fig:attack1_1}
\end{figure}

\textit{Concealment:} The actual malicious logic has been hidden
inside an Add-On Instruction. A new instruction has been created,
which is very similar in its construction to the real \emph{ADD}
block, with similar inputs: 2 sources A and B and an output:
Destination. It has also been named suitably (\emph{ADD\_A}) to
disguise well with a real \emph{ADD} block. From the top overview of
the ladder logic (which contains many rungs), this looks just like any
other \emph{ADD} block on one of the rungs. But inside this add-on
instruction, the real bomb (an infinite loop) is defined, and that
adversely affects the PLC operation. More details about this can be
found in Figure~\ref{fig:attack1_1}.
\subsection{Attack 2: Manipulation of Sensor data using Subroutines}
Another important function of the PLCs in ICS (in addition to
controlling the actuators) is reading data from sensors. That data can
be critical information about the process and system. Using the data,
it is possible to derive the current state of the process, which is
used by the PLC to determine appropriate control actions. Thus, tampering with   sensor data can cause systems to fail~\cite{kosut10malicious}.

\textit{Goal:} The goal for this attack was to manipulate sensor readings
coming from the remote IOs (RIOs in Figure~\ref{fig:generic}) to the PLC.

\begin{figure}[tb]
\centering
\includegraphics[width=\linewidth]{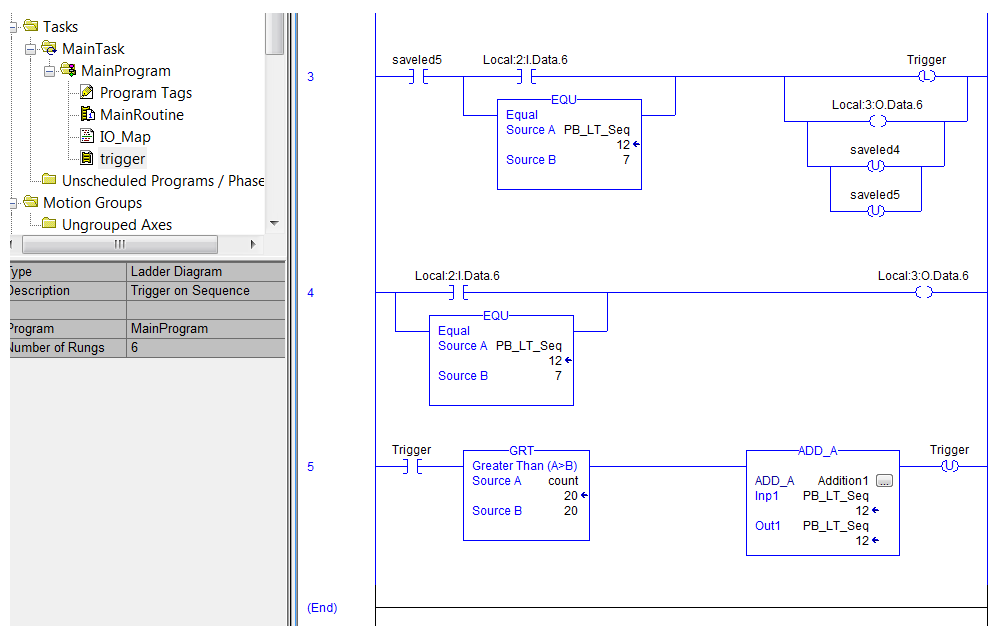}
\caption{Inside the exploiting subroutine}
 \label{fig:attack2_1}
\end{figure}

\textit{Construction:} Since this is proof-of-concept, we decided to
manipulate the sensor values and increase them by a constant offset
(we arbitrarily chose four). As result, the LLB payload is a simple
\emph{ADD} block which takes the real sensor values and increases them
by four, and stores them back into the same tag.  However, a more
complex triggering mechanism was used in this attack. In particular,
the LLB is triggered when a complete trigger sequence is
detected. This has been achieved by implementing a finite state
machine using latches (see Figure~\ref{fig:attack2_1}).

\begin{figure}[tb]
\centering
\includegraphics[width=\linewidth]{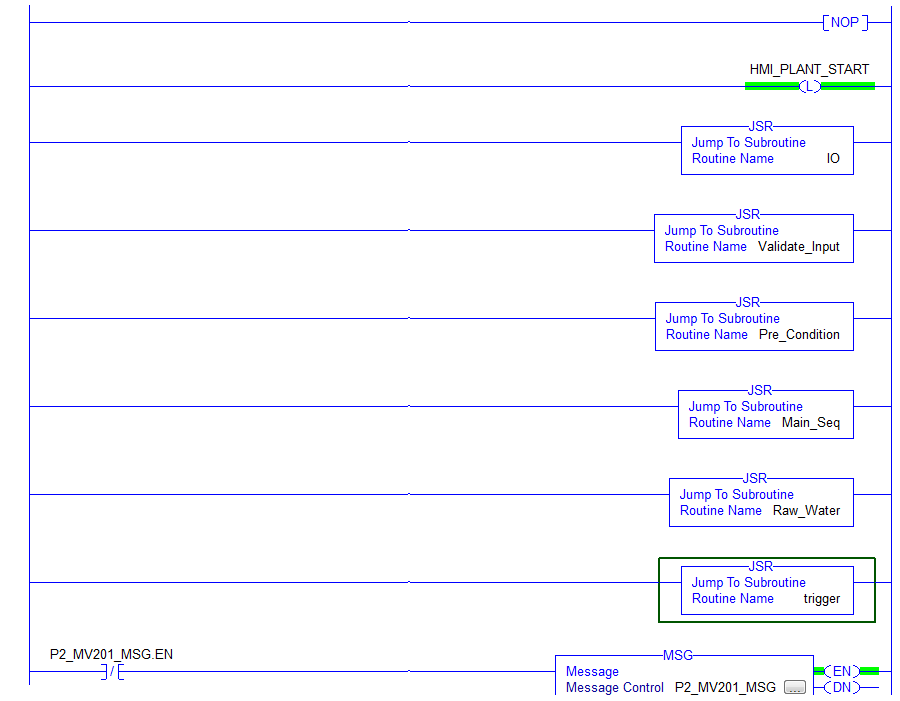}
\caption{Overview of the logic with the exploiting subroutine}
 \label{fig:attack2_2}
\end{figure}

\textit{Concealment:} For this attack, we also used A different hiding
technique. By inspecting the actual logic of the PLC in the water
treatment plant, we observed that the logic was calling a large number
of subroutines. We assume the subroutines were called that way to
maintain good readability of the ladder logic by the
maintainers. However, that structure with large number of subroutines
can be leveraged by the attacker to hide the LLB. We tested this
exploit by hiding a trigger subroutine that gets executed every cycle
of the ladder logic (see Figure~\ref{fig:attack2_2}).
\subsection{Attack 3: Data Logging using FFLs} 
The attacks discussed above are openly causing damage or malfunctions,
and their effects can be observed as soon as triggered. However, there
are another class of LLBs which can be equally harmful but are harder
to detect. In particular, such LLBs could be used for data logging and
exporting sensitive information about the system.

\textit{Goal:} The goal of this attack is to achieve stealthy data
logging of sensitive information about the plant.

\begin{figure}[tb]
\centering
\includegraphics[width=\linewidth]{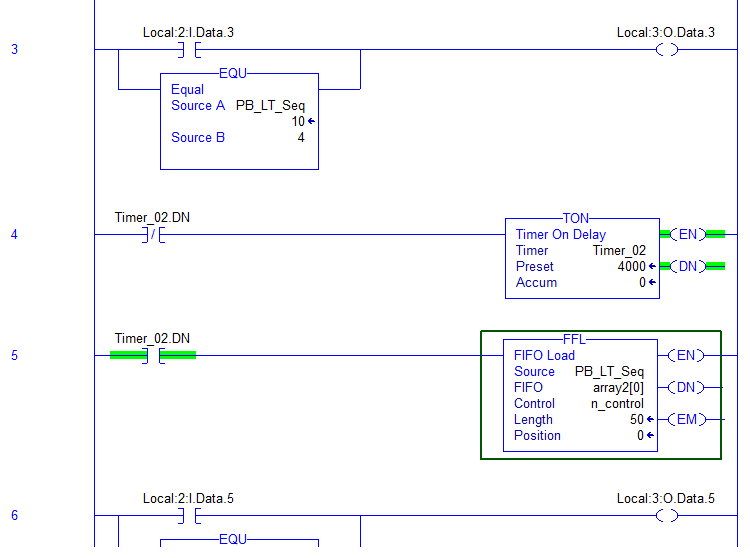}
\caption{Data logging in a FIFO buffer}
 \label{fig:attack3_1}
\end{figure}

\textit{Construction:} The data logging is achieved by using a FIFO
buffer which reads data into an array. The \emph{FFL} block has been
used for this purpose. As shown in Figure~\ref{fig:attack3_1}, the
\emph{FFL} block stores the tag \emph{PB\_LT\_Seq} which contains
sensitive information about the count sequence used to determine state
of the plant. Those values are stored into the \emph{array2} and are
converted into \emph{.csv} format and stored on the SD card in the
PLC. Staying within our attacker model, an attacker who has sporadic
access (physical access to PLCs) to the plant can come in, read these
values stored on the SD card. Then, insert this card back into the PLC
and leave. The trigger sequence for this could be a simple timer, thus
ensuring data logging after 'x' days of plant operation.

\textit{Concealment:} This LLB can again be concealed either inside an
Add-On instruction or as a subroutine. It can also be left inside the main
logic flow, since this LLB contains just one extra rung, making its
manual detection difficult in large and complex code.
\subsection{Attack 4: Trigger Major Faults on PLC} 
We now discuss another attack which is similar in effect to the DoS
attack.

\textit{Goal:} The goal is to trigger major faults on the PLC which
causes its processor to halt and which cannot be fixed by a hard
reset.

\begin{figure*}[tb]
\centering
\includegraphics[width=\linewidth]{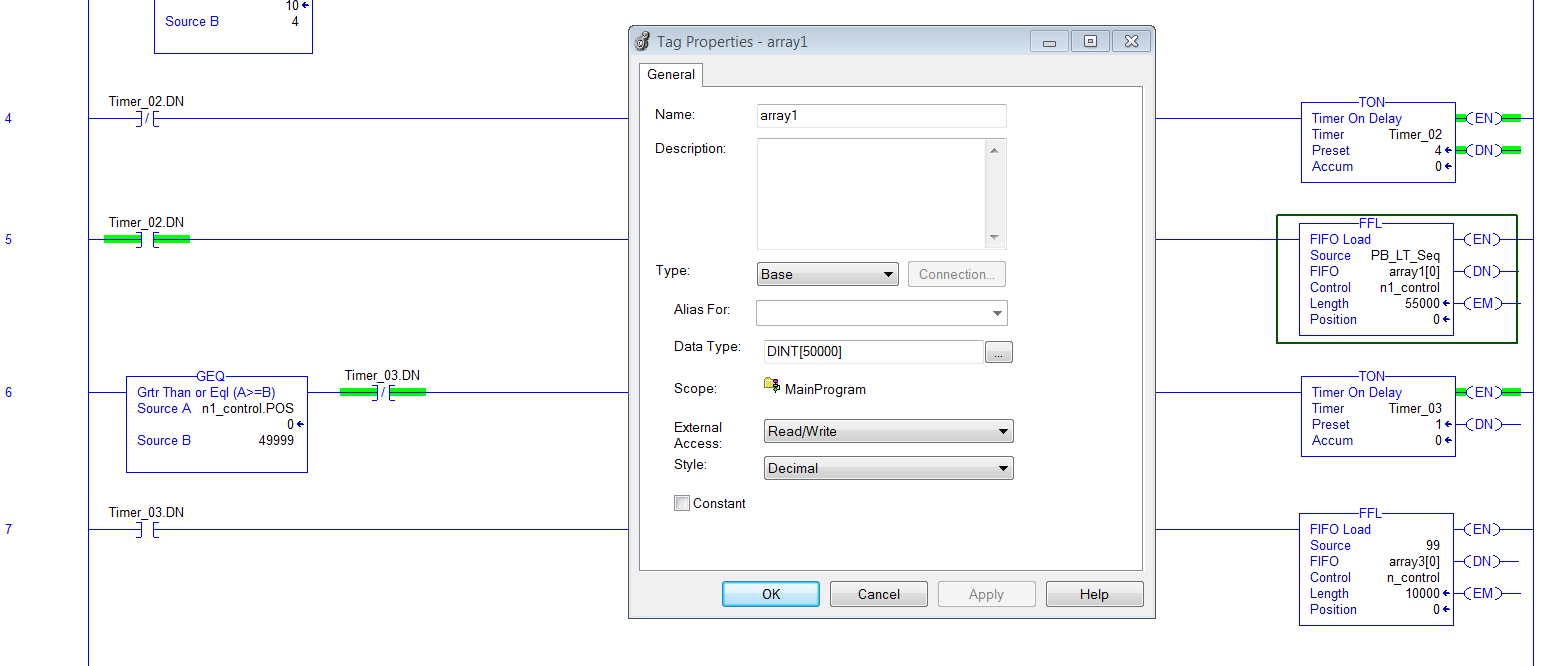}
\caption{Invalid Array Subscript}
 \label{fig:attack4_1}
\end{figure*}

\begin{figure}[tb]
\centering
\includegraphics[width=\linewidth]{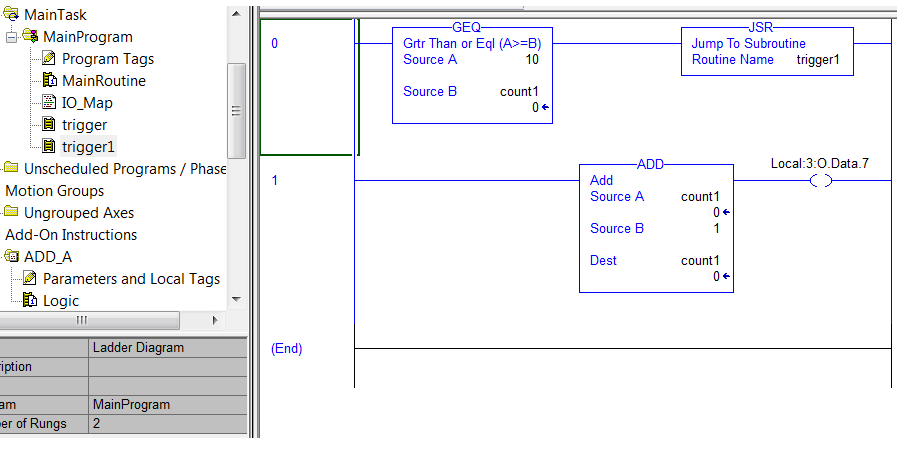}
\caption{Stack Overflow}
 \label{fig:attack4_2}
\end{figure}

\textit{Construction:} Here we managed to cause two major faults on the PLC.
\begin{enumerate}
\item Invalid Array Subscript \\
This was achieved by causing an overflow in the array used for collecting
tag information. This can be done by creating a mismatch between the FIFO
buffer length and size of the array used to store values of the buffer.
Details can be found in Figure~\ref{fig:attack4_1}.
\item Stack Overflow \\
This was achieved by implementing a recursive subroutine call to itself.
This caused the stack storing the return pointer to overflow, halting the
process and crashing the PLC (Figure~\ref{fig:attack4_2}).
\end{enumerate}
\textit{Concealment:} These LLBs can be concealed within an Add-On instruction or inside a subroutine. 


\subsection{Analysis of Attacks}
Ideally, there would be a metric to measure the \emph{stealthiness} of
LLBs, that would indicate how hard different LLBs are to discover. So
far, we have not found a good way to measure that property. In the
following, we instead use the \emph{relative additional lines of code}
(RALOC) to measure the stealthiness. In particular, the increase of
lines of code in the logic can also lead to increased memory
consumption at runtime. We observed that there are two types of memory
that is used by a ladder logic program: I/O memory and Data \& Logic
memory.  As part of our analysis, we measured the difference
(increase) in memory of the original logic when malicious ladder logic
bombs were added.  It was observed that there was no increase in the
I/O memory of the PLC at all, which is primarily because no new
inputs/outputs were created to trigger or apply the ladder logic bombs
discussed above. The only increase observed was in the data and logic
memory, which is also marginal, as depicted in
Table~\ref{tab:comparison}. One important thing to note is that the
size of Attack 3 (data logging) will depend on the amount of data
that is logged. As result, the RALOC metric increases, and the
modifications might become more visible.

To mitigate that effect, it is best to save the data on the SD card
and then flush the arrays so that they can be re-used if more data
needs to be logged.
\begin{table} 
\caption{Comparison of Attacks Performed}
\label{tab:comparison}
\begin{tabular}{lc} 
Attack&Increase in Memory(\%)\\ \toprule
Attack 1: DoS using AOI&2.60\\ 
Attack 2: Manipulate Sensor&3.84\\ 
Attack 3: Data Logging&3.41\\ 
Attack 4: Major Faults&4.09\\ \bottomrule
\end{tabular}
\end{table}

\section{Evaluation}
\label{sec:evaluation}

\subsection{Evaluation Context}
\label{sec:context}

To estimate the difficulty for humans to detect LLBs, we ran a
small-scale challenge as part of an event organized at our
institution. Six teams from academia and industry participated in the
event, and received three challenges related to LLBs. We note that not
all participants were very familiar with ladder logic programming, but
each team was provided a testbed manual to understand the
overall setup, software use, and tag initialization.

The challenges were run remotely with teams in different locations
around the world, connection to a virtual operator machine and a
physical PLC in our lab. In particular, the virtual machine was
configured with Studio 5000 and RSLinx to provide communication to the
testbed PLC. The participants connected to the virtual operator
machine through a virtual private network (VPN).

For all three challenges, the PLCs were programmed with a basic
configuration to interact with the IOs and send selected control
signals.  We now summarize the challenges, which involved a brief
description of the problem statement, along with specific goals to
achieve.

\paragraph{Jump to catch the flag} The first challenge goal was to
detect an LLB that was designed to read a connected sensor,
manipulate that reading, and then potentially forward that reading to
the outside world. To solve that challenge, the participants had to
follow the data flow coming from all sensors to the control functions,
and identify parts of the code that should not need the sensor value,
but used it nevertheless. After identifying the LLB code, the
participants could then read the created tag value to obtain the flag.

\paragraph{Play with Add\_on Instructions} The second Challenge was to
get the true value of a connected analog sensor that was read by some
logic. To solve the challenge, the participant had to detect an LLB
that tried to hide as an Add\_on instruction (see
Figure~\ref{fig:attack1_1}). Once the participant detected the LLB,
they were able to remove it to obtain the true value (or simply determine the applied offset, and remove it manually).

\paragraph{Fix Me if you can} The third challenge consisted of logic
that contained a programming error. When run on the PLC, the code
would lead to a ``PLC Major Fault'' error message, and stop
executing. In particular, we wrote the code to access a memory array
with an index that exceeded the length of the memory array (similar to
a buffer overflow).  Such a fault could be used as LLB payload to shut
down operations of a PLC. To solve that challenge, the participants
had to understand the FFL block and detect that an uninitialized
memory access can lead PLC to faulty state.

\subsection{Challenges Results}
\label{sec:ctf-res}

This section summarized the challenge results as obtained during the
CTF event. The details of the teams are anonymized. One of the team is
not included in the analysis, as the team managed to obtain the flag
through an unrelated side-channel. The other teams' results are
summarized in Table~\ref{tab:eval}. Only Team 2 was able to solve all
the challenges (i.e., they were able to detect all the LLBs), and Team
5 was able to solve one challenge. The remaining teams were not able
to detect any LLBs.

\begin{table}[h]
\centering
\caption{LLB Evaluation details.\label{tab:eval}}
\begin{tabular}
{l | @{}c@{} @{}c@{} @{}c@{} }
\toprule

Teams & {\ First Bomb \ } & {\ \ Second Bomb} & {\ \ Third Bomb} \\
\midrule

Team 1            & \Tdot & \Tdot & \Tdot\\
Team 2            & \Twdot & \Twdot & \Twdot\\
Team 3            & \Tdot & \Tdot & \Tdot\\
Team 4            & \Tdot & \Tdot & \Tdot\\
Team 5            & \Twdot & \Tdot & \Tdot\\

\bottomrule
\end{tabular}\\
\footnotesize{
    Legend\quad
    \Twdot:   Detected,\quad
    \Tdot:  Undetected.
}
\end{table}

Our (limited) evaluation shows that detecting malicious code or hidden
logic bombs in critical infrastructure controller code is not a
trivial task.  Only two teams were able to find the LLBs among a large
number of subroutine calls along with several message and instruction
blocks. The more advanced challenges which included the LLB hiding as
Add\_on instruction were only solved by one team. We conclude that in
order to detect LLBs, an operator must have sound knowledge of Studio
5000 and programming languages like ladderlogic, Structure text, and
functional block diagram along with its syntactical and semantic
meaning. In practice, that can be challenging if an operator has to
inspect code with ill-specified functionality or written by a
subcontractor.

\section{Countermeasures}
\label{sec:countermeasures}

In this section, we discuss potential countermeasures against LLB
attacks. In particular, we discuss a) network-based countermeasures,
and b) centralized validation of running code.

In the following, we assume that the countermeasures are retro-fitted
into an existing industrial control system. In particular, we assume
it is not possible to change the PLCs themselves. If we could change
the way logic updates are applied to PLCs, it would trivially be
possible to introduce user authentication (e.g. with
username/password, or public key-based), or cryptographic signatures
for logic updates. The PLC would then only accept the logic code
update if the user is successfully authenticated, or the authenticity
of the update has been validated.

The following two proposals do not require such changes to the
existing PLCs, and should thus be easier to implement in existing
systems. In the following, we assume that there are a number of
well-known operators in the ICS, that are allowed to update the
control logic of the PLCs. Any attempts to update PLC logic by other
third parties is counted as an attack. We assume that the default
software is used to apply logic updates (e.g., Studio 5000), and that
we cannot change the behaviour of that software (e.g., we cannot add
additional authentication information into traffic generated by it).

We assume the attacker model from before: the attacker has the
capability to manipulate the logic running on a PLC once, but does not
have permanent access. The attacker did not compromise the operator's
machine. The attacker is also not able to manipulate third party
network traffic.

\subsection{Network-based countermeasures}
If an intrusion detection system (IDS) is already used in the network
to monitor traffic for spreading malware or other malicious traffic,
then that IDS could potentially be used to identify the specific
traffic related to logic updates on PLCs connected to the network. If
unauthorized logic updates over the network are observed, an alarm
could be raised.  A similar IDS is proposed
in~\cite{goldenberg2013accurate}, where the authors model periodic
communication between HMI and PLCs using a deterministic finite
automata. The system flags anomalies if a message appears out of
position in normal (general) sequence of messages.  If the IDS is
configured to operate as intrusion prevention system (IPS), the
offending traffic could even be dropped in real time.

The problem with this proposal is related to the identification of
authorized logic updates. As we cannot change the traffic generated by
the respective software, there is no way to embed specific
authentication information. Thus, we can only use information such as
IP source address (supposedly related to the authorized person), which
is not ideal (as it can be spoofed).

\subsection{Centralized Logic Store}
\begin{figure}[tb]
\centering
\includegraphics[width=\linewidth]{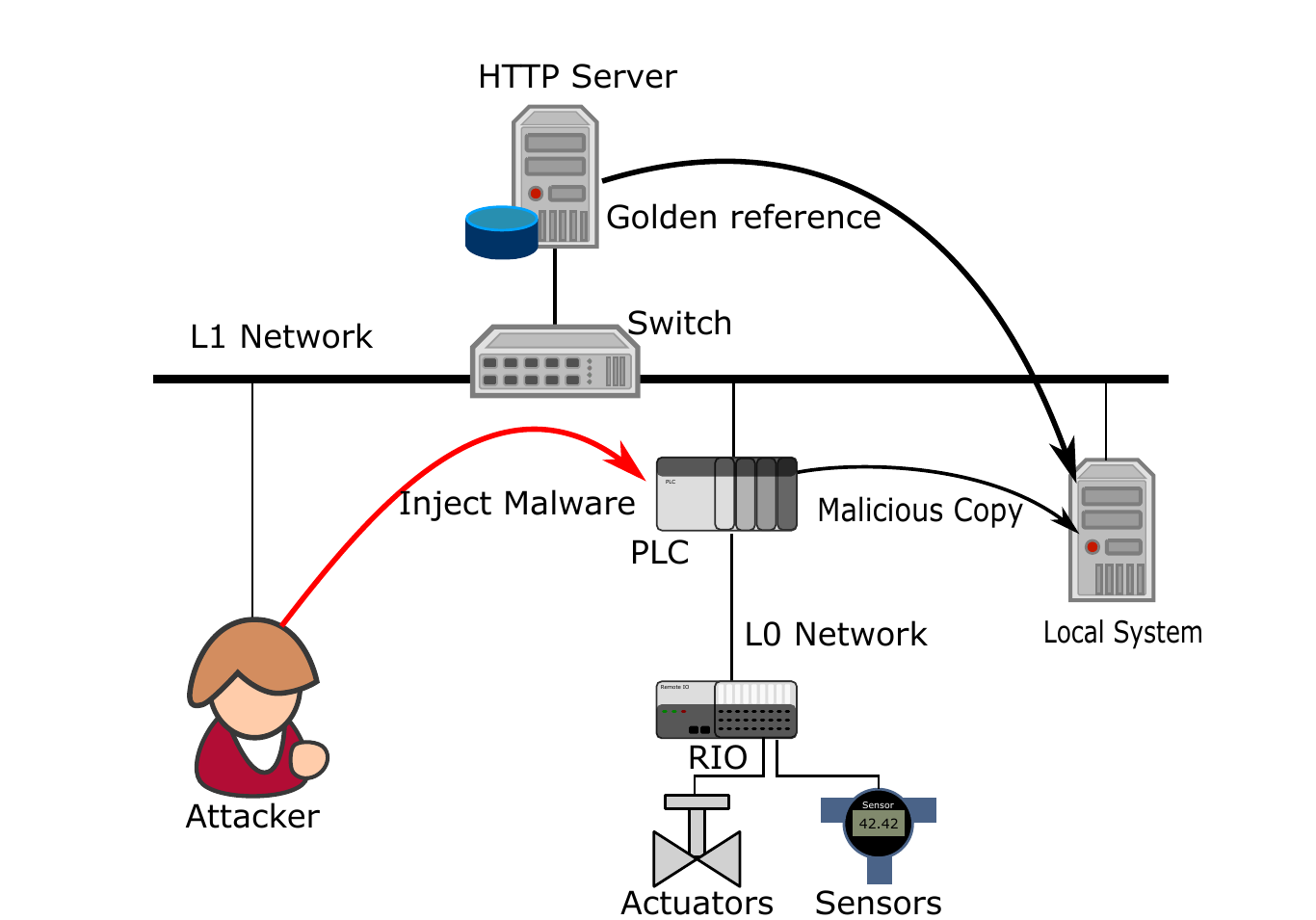}
\caption{Centralized Logic Store based countermeasure}
 \label{fig:cslfig}
\end{figure}

Our second proposal is based on two components: a) a centralized logic
store (CLS) of the latest version of logic running on all PLCs of the
ICS, and b) a tool to periodically download currently running logic
from the PLCs, and to validate that against the ``golden'' copy from
the CLS.  An overview of our proposed system can be found in
Figure~\ref{fig:cslfig}.

\paragraph{Submission of golden samples}
All authorized engineers are required to submit the most recent
version of logic for each PLC to the CLS when they change the logic
running on the PLCs. To do so, they can use a simple application that
requires them to identify the respective logic file, the target PLC,
and their credentials. That application will then use the credentials
to establish an authenticated secure channel to the CLS (e.g. using
TLS), and then upload the latest logic version to the CLS (e.g. using
HTTP over the established TLS session).

\paragraph{Periodic Logic Validation} We have implemented a
python-based tool to manually and periodically validate the logic. The
user first exports ladder logic to a \emph{.L5K} file (sequential
text) on the local machine using Studio 5000. Next, our tool parses
the \emph{.L5K} file and extracts a unique serial number corresponding
to the logic. Then, the tool connects to the CLS where the correct
golden logic is searched by using Beautiful Soup parser (BSP). BSP is
a python library to parse HTML and XML pages, in our case BSP parse
CLS and look for all \emph{.L5K} file followed by our parser which
looks for correct golden logic by identifying the unique serial
number.

Then, the tool performs a comparison between the logic found on the
PLC, and the golden sample. If differences are found, they can be
visualized to a human operator using standard functionality provided
by tools such as \emph{diff}. The algorithm below summarizes the whole
process.

\begin{algorithm}
\caption{CLS based countermeasure}
\begin{algorithmic}
\REQUIRE Downloaded malicious PLC logic (.L5K) file\\
 Establish server connection at specific port
\STATE Parse local .L5K file and fetch serial no. 
\STATE GET golden sample from server with serial no.
\IF{diff(local .L5K,golden reference .L5K) == 0}
\STATE Local logic successfully validated
\ELSE
\STATE Local logic differs, present diff to user
\STATE User manually inspects code differences
\IF{User detects attack}
\STATE Raise alarm
\ELSIF{Local Logic newer}
\STATE Update golden sample on CLS if authorized
\ELSE
\STATE Update local logic with golden sample
\ENDIF
\ENDIF
\end{algorithmic}
\end{algorithm}

There is a tool developed by Rockwell Automation called Factory Talk
AssetCentre, which tries to achieve similar functionality in securing
PLC devices. However it has many additional dependencies, for example:
need for a network adapter card on both client/server side,
FactoryTalk services platform, RSLinx, RSLogix 5000, etc. The proposed CLS 
based approach that we have described above is easy to use, in contrast 
to Factory Talk AssetCenter which requires a operator, having sound knowledge of system 
requirements and capacity. The CLS based approach is much more complete,
dependency-free and general purpose to use across platforms/PLCs from
different vendors. 

\section{Related work}
\label{sec:related}

\Paragraph{General Threats to ICS}
It has been observed over the years that process control systems are
vulnerable to various exploits with potentially damaging physical
consequences~\cite{cardenas2008research,liu2011false,amin2010stealthy,pollet2010scada}.

In~\cite{morris13cyberattack} Morris \emph{et al} discuss different
attacks such as measurement injection, command injection, denial of
service, etc, on SCADA control systems which use the MODBUS
communication protocol. Much like the rest, this study is again
restricted to exploiting the network layer to attack the
PLCs. Therefore, it is necessary to analyze control logic
vulnerabilities, which can be manifested through malicious logic
additions.

The authors of the cited related generally highlight that ICS/SCADA
systems are threatened by attacks, despite wide-spread use of \emph{air
  gaps} between the Internet and ICS network. 

\Paragraph{Stuxnet}
In 2010, Stuxnet~\cite{falliere2011w32} caused a radical shift in
focus for security of such control systems by demonstrating practical
exploitation of the control logic in these devices. This resulted in
increasing focus on security aspects of PLCs and their control
logic~\cite{beresford2011plc,karnouskos2011stuxnet,kim2014cyber}.

In ~\cite{karnouskos2011stuxnet}, Karnouskos \emph{et al.} discuss
Stuxnet and how it managed to deviate the expected behaviour of PLC.
In~\cite{kim2014cyber}, Kim \emph{et al.}  discuss the
  cyber security issues in nuclear power plants and focused on stuxnet 
inherited malware attacks on control system, and its impacts in future
along with its countermeasures.

\Paragraph{Protocol-based attacks} The authors
of~\cite{beresford2011plc} discuss replay, reconnaissance and
authentication by-pass attacks. These attacks can be performed by
sending probe requests or by examining the ISO-TSAP conversation and
authenticating oneself by generating packets with same hash, in turn,
achieving access to PLC logic. All these attacks are focused on
exploiting the communication protocols to gain access to PLCs. 

In~\cite{milinkovic2012industrial}, the authors investigate
vulnerabilities of industrial PLCs on firmware and network level,
leaving out any analysis on logic level exploits. In this work, we
provide a consolidated study on logic layer manipulations and provide
logic level safeguarding methods, unlike the network based security
(e.g., firewall, VPN security and secured layered architecture)
methods proposed in majority of the papers above.

\Paragraph{Control Logic Manipulation} In~\cite{mclaughlin11hotsec}, the authors propose a PLC malware capable
of dynamically generating a payload based on observations of the
process taken from inside the control system. The malware first
gathers clues about the nature of the process and the layout of
physical plant. Dynamic payload is then generated to meet the specific
payload goal. However, the authors assume that an attacker must be
insider or have prior knowledge of the targeted system. That
dependency is worked upon in~\cite{mcLaughlin12specification}, which
proposes a tool to automatically determine semantics of the target
PLC, minimizing the need for prerequisite knowledge of target control
system. This work however does not go into details of malicious logic
construction on ladder logic or any other IEC 61131-3-compatible
language and focus mainly on network layer attack.

\Paragraph{Countermeasures}
In general, attempting to validate the authenticity of the root file
system or files/directories is not a new concept. 
In~\cite{kim1994design}, Kim et al. proposed a monitoring tool
"Tripwire". It monitors the Unix based file system and notifies the
system administrator in case a corrupted file or alteration is
detected. In contrast to Tripwire tool (which uses interchangeable
signature subroutines to identify changes in file) our proposed CLS
based countermeasure compares the local instance of a file with its
authorized one. Another important point to note here is that the
Tripwire tool is host based, used for unix based file systems whereas
proposed countermeasure is used in respect to PLC logic file(.L5K)
extracted from Studio 5000 tool.

We found a number of works focused on development of countermeasure
techniques to safeguard PLCs and other components of industrial
control systems. In~\cite{caselli2015sequence}, a sequence aware
intrusion detection system (S-IDS) is proposed. The IDS focuses on
detection certain sequences of events (e.g. sensor readings or control
actions) that are harmless on their own, but can lead to unwanted
consequences if chained
together. In~\cite{hadvziosmanovic2014through}, the authors propose a
detector which monitors process variables continuously to ascertain
changes and attacks. Other attack detection methods for PLCs are found
in~\cite{zonouz2014detecting} and~\cite{mclaughlin2014trusted}.
In~\cite{zonouz2014detecting}, the authors propose an approach based
on symbolic execution of PLC code along with control model checking to
automatically detect the malicious code running on the
PLC. In~\cite{mclaughlin2014trusted}, a Trusted Safety Verifier (TSV)
is implemented on a Raspberry PI set-up, placed in between the control
system network and the PLC as a bump-in-the-wire to intercepts all the
controller code and validate it against all the safety properties
defined by process engineer. This requires additional hardware set-up
to function. In this paper, we intend to propose countermeasures which
can be very easily used with the traditional (existing) industrial
control system architecture and have least dependency on PLC internals
(construction and interface internals).

\section{Conclusion}
\label{sec:conclusions}
In this paper, we have introduced the term \emph{ladder logic bombs}
to discuss the problem of logic malware for PLCs, such as
modifications performed by Stuxnet~\cite{falliere2011w32}.
Contemporary vulnerabilities study for such devices usually do not
include analysis on control logic level, which is an important source
of attacks as demonstrated in this work. We analyzed vulnerabilities
in the firmware running on PLCs and depicted case studies and attack
scenarios in real-time on actual PLCs to inflict damage on industrial
control systems. Through a small-scale evaluation, we have shown that
even simple LLBs can be hard to detect in real-world control logic
code.  All the tests were conducted on a real world ICS, unlike
majority of the theoretical works presented in the literature so
far. Finally, a centralized logic store based countermeasure technique
was proposed and implemented, that can detect logic level based
attacks effectively.

\section{Acknowledgments}
We thank Nicolas Iooss for his support and contributions related to
EtherNet/IP support in MiniCPS and the demonstrated attacks, and
Pierre Gaulon for his help on the physical layer simulation. This work
was partially supported by SUTD's startup grant SRIS14081.

\balance 
\bibliographystyle{abbrv}
\bibliography{bibliography}

\end{document}